\documentclass[runningheads]{llncs}
\usepackage[T1]{fontenc}
\usepackage{graphicx}
\begin{document}
\title{A Cost-Effective Edge Computing Gateway for Smart Buildings}
\titlerunning{A Cost-effective Edge Computing Gateway for Smart Buildings}
%
%
\author{Simon Soele Madsen\inst{1}\orcidID{0009-0008-5279-9500} \and
Benjamin Eichler Staugaard\inst{1}\orcidID{0009-0007-3392-4567} \and
Zheng Ma\inst{1}\orcidID{0000-0002-9134-1032} \and
Salman Yussof \inst{2}\orcidID{0000-0002-2040-4454} \and
Bo Nørregaard Jørgensen\inst{1}\orcidID{0000-0001-5678-6602}
}
\authorrunning{S. Madsen et al.}
%
\institute{Center for Energy Informatics, Maersk Mc-Kinney Moller Institute, University of Southern Denmark, 5230 Odense, Denmark \and Institute of Informatics and Computing in Energy, Universiti Tenaga Nasional, Kajang 43000, Malaysia \\
\email{ssma@mmmi.sdu.dk, beis@mmmi.sdu.dk, zma@mmmi.sdu.dk, salman@uniten.edu.my, bnj@mmmi.sdu.dk}}
\maketitle              
\begin{abstract}
The retrofitting of existing buildings with building management systems presents significant challenges, primarily due to the need for labor and cost efficiency. Wireless technology offers a promising solution to these challenges by minimizing the need for extensive wiring and structural alterations. However, achieving retrofitting in a cost-effective manner necessitates the use of low-cost wireless technologies. This paper introduces a framework for constructing a Zigbee gateway using open-source tools combined with low-cost hardware. The proposed architecture addresses large-scale IoT deployments within the Zigbee ecosystem. By leveraging edge computing with the robustness and scalability offered by Zigbee technology, this architecture significantly reduces the economic barriers to retrofit buildings with building management systems. The results underscore the potential of open-source Zigbee technology in aligning with sustainability goals, providing a cost-effective pathway for retrofitting buildings into smart, energy-efficient living environments.

\end{abstract}
\section{Introduction}
Today buildings are responsible for 28\% of global energy consumption~\cite{iea}, and the European Union deems that around 60\% of the total building stocks are poor energy performers~\cite{eu}. To address the pressing need to mitigate the adverse effects of energy inefficient buildings and their contribution to climate change, policymakers worldwide have enacted stringent directives and regulations aimed at fostering the emergence of greener buildings. The European Union has introduced the Fit for 55 package~\cite{fitfor55} and the Energy Performance of Buildings Directive (EPBD)~\cite{epbd}. These initiatives aim to significantly reduce greenhouse gas emissions and improve energy efficiency across various sectors, including buildings. The EPBD is a legislative framework aimed at improving the energy performance of buildings within the EU. It aims to have a zero-emission and fully decarbonised building stock by 2050. The EPBD highlights the integration of smart technologies to create smart buildings, enabling real-time energy system monitoring and control, optimized heating, cooling, and lighting, and predictive maintenance.

Given that a significant portion of the current building stock requires retro-fitting to meet these stringent standards ~\cite{epbd}, the adoption of wireless technologies presents a highly effective solution. Wireless technologies offer several advantages for retrofitting existing buildings into energy-efficient smart buildings~\cite{Madsen2021,Ma2022}. These technologies are minimally invasive, eliminating the need for extensive wiring and structural alterations, making the retrofitting process faster, less costly, and less disruptive to building occupants. Zigbee~\cite{csa-zigbee}, in particular, stands out due to its low-power wireless communication protocol and cost-effectiveness. Its ability to form mesh networks ensures robust connectivity and extended range, which is crucial for covering large building areas. However, despite its potential, the Zigbee ecosystem faces challenges related to intrinsic interoperability issues, which can hinder its widespread adoption and integration into smart building ecosystems.

This paper aims to explore the role of open-source Zigbee technologies in the transformation of existing buildings into energy-efficient smart buildings. By examining the challenges posed by interoperability issues within the Zigbee ecosystem and the imperatives outlined in new building directives~\cite{epbd}, this study seeks to highlight how open-source Zigbee technologies can align with sustainability objectives. Through an analysis of existing literature, a case study, and technological advancements, this paper describes the mechanisms through which open-source Zigbee can empower stakeholders in the building sector to realize energy efficiency and thus sustainability in a cost-effective way.

This paper is structured as follows: The literature review section presents a comprehensive comparison of wireless technologies. The methodology section outlines the architecture and implementation of the open-source Zigbee gateway, detailing the hardware and software components. The case study section demonstrates the practical application of the proposed gateway in a real-world smart building environment. The results section presents the case study findings, discussing performance metrics. The discussion section addresses interoperability challenges and explores potential cost reductions and customizations. The conclusion summarizes the key contributions, emphasizing the solution's efficiency, reliability, and sustainability.

\section{Literature Review}
When retrofitting buildings with building management systems using wireless technologies, it is crucial to compare the equipment cost, signal range, and battery lifetime of alternative technologies to ensure both economic feasibility and long-term operational efficiency. Table~\ref{tab:wireless} provides a comparison of various wireless technologies, note that the actual power consumption for Zigbee, Z-Wave, and 6LoWPAN is drastically lower due to the sleepy nature of end devices. The results are from comparison studies~\cite{Tsantilas2020,9374808} and technical specifications~\cite{IEEE802.11ac,ZWaveSpec}. Each technology presents distinct advantages and limitations in terms of frequency band, data transfer rate, signal range, power consumption, mesh topology, and market availability.

\begin{table}[]
\caption{Wireless technology comparisons.}
\label{tab:wireless}
\resizebox{\textwidth}{!}{\begin{tabular}{|l|l|l|l|l|l|}
\hline
\multicolumn{1}{|c|}{\textbf{Feature}}                                & \multicolumn{1}{c|}{\textbf{WiFi}} & \textbf{Zigbee} & \textbf{Z-Wave} & \textbf{BLE} & \textbf{6LoWPAN} \\ \hline
\textbf{\begin{tabular}[c]{@{}l@{}}Frequency \\ Band\end{tabular}}    & 2.4/5 GHz                          & 2.4 GHz         & 868/915 Mhz     & 2.4 GHz      & 2.4 GHz          \\ \hline
\textbf{\begin{tabular}[c]{@{}l@{}}Data\\  Transfer\end{tabular}}     & Up to 1.73 Gbps                    & Up to 250 Kbps  & Up to 100 Kbps  & Up to 2 Mbps & Up to 250 Kbps   \\ \hline
\textbf{Range}                                                        & Up to 100m                         & Up to 100m      & Up to 100m      & Up to 10m    & Up to 100m       \\ \hline
\textbf{\begin{tabular}[c]{@{}l@{}}Power \\ Consumption\end{tabular}} & 250 mW                             & 250 mW          & 250 mW          & 10 mW        & 250 mW          \\ \hline
\textbf{\begin{tabular}[c]{@{}l@{}}Mesh\\ Topology\end{tabular}}      & No                                 & Yes             & Yes             & Yes          & Yes              \\ \hline
\end{tabular}}
\end{table}

WiFi, operating in the 2.4 GHz and 5 GHz bands, offers high data rates up to 1.73 Gbps, making it suitable for high-bandwidth applications like video streaming. However, its high-power consumption (250 mW) and lack of mesh networking limit its scalability and efficiency for battery-powered sensor networks in buildings. In contrast, Zigbee, which also operates in the 2.4 GHz band, provides lower data rates up to 250 Kbps but excels with low power consumption (250 mW, typically much lower in practice) and mesh networking support, making it ideal for large-scale sensor deployments.

Similarly, Z-Wave, operating at 868/915 MHz, offers data rates up to 100 Kbps and supports mesh networking with a range of up to 100 meters. Although it shares Zigbee’s low power consumption and mesh capabilities, Z-Wave suffers from limited market availability, making device selection and deployment more complex compared to Zigbee’s robust market presence.

Bluetooth Low Energy (BLE), operating in the 2.4 GHz band, combines higher data rates (up to 2 Mbps) with very low power consumption (10 mW) and supports mesh networking. However, its limited range of up to 10 meters restricts its utility in building-wide sensor networks, unlike Zigbee and Z-Wave, which offer more extensive coverage.

6LoWPAN, compatible with both 2.4 GHz and sub-1 GHz bands, offers data rates up to 250 Kbps and supports mesh networking with low power consumption (250 mW, typically much lower in practice). Despite these advantages, its lower market availability compared to Zigbee poses challenges for integration in widespread building retrofits.

In summary, while WiFi is suitable for high-bandwidth needs, its high-power consumption and lack of mesh networking make it less ideal for dense sensor networks. Zigbee stands out with its low power consumption, mesh networking, and strong market presence, making it highly suitable for sensor deployments. Z-Wave offers similar benefits but is hindered by limited market availability. BLE’s very low power consumption is advantageous, but its short-range limits building-wide use. 6LoWPAN, with its mesh networking and low power consumption, is effective for specific applications but less favorable for broad deployment due to market constraints.

The cost-effectiveness of devices is an important factor when retrofitting buildings. This consideration impacts not only the initial investment but also the long-term operational costs associated with maintaining and upgrading the infrastructure. Lower-cost devices with efficient energy consumption can significantly reduce overall expenses, making smart building technology more accessible and sustainable. Additionally, cost-effective solutions often allow for more extensive sensor deployment, enhancing the building's functionality and efficiency. Therefore, selecting affordable yet reliable technologies ensure a higher return on investment and encourages widespread adoption of smart building innovations.

For example, comparing the prices of CO2 sensors across different technologies illustrates the potential cost savings. A CO2 sensor utilizing 6LoWPAN technology, which also measures temperature and humidity, costs 250€~\cite{LafipaCO2Sensor}. This sensor requires a power supply during the initial calibration phase. In contrast, a similar Zigbee CO2 sensor, which also measures temperature and humidity, is available for just 99€~\cite{NexelecCO2Sensor}. The significant price difference highlights Zigbee's cost-effectiveness, making it a more attractive option for extensive sensor deployments in smart building retrofits. Therefore, from here on out, we shift our focus to Zigbee for building management systems.

Zigbee has been widely used in energy management systems within smart buildings to monitor and control energy consumption. For instance, a Zigbee-based building energy monitoring and control system can effectively monitor energy usage with high accuracy, facilitating long-term energy conservation planning and the development of automated energy conservation strategies for building applications~\cite{Peng2014,Peng20161615}. Similarly, a smart home energy management system that utilizes Zigbee for energy measurement modules can monitor the energy consumption of home appliances and integrate renewable energy sources via power line communication gateways to optimize home energy use and reduce energy costs by considering both consumption and generation data~\cite{Han2014198}.

In another study, home automation systems demonstrated a significant potential for power management and energy conservation, with findings showing an 18.70\% decrease in energy consumption in homes equipped with such systems~\cite{Tejani2011241}. This highlights the benefits of smart home technology in promoting energy efficiency and sustainability.

To address compatibility challenges, solutions such as using separate gateways/bridges for each network and integrating them with an open-source home automation framework like OpenHAB have been proposed. This approach allows for a single point of control through an application or web interface, improving the user experience and addressing vendor lock-in issues~\cite{Chaudhary20212367}.

Smart buildings leverage Zigbee-based systems for various applications, including environmental monitoring, energy management, and automation. These systems provide significant benefits, such as improved energy efficiency, enhanced occupant comfort, and reduced operational costs. A cost-effective and scalable sensor network for intelligent building monitoring utilizing Zigbee devices can integrate multiple sensors for real-time monitoring and smart building management through autonomous software agents, helping to identify wasted energy consumption and suggest better usage of building spaces~\cite{Dibley20128415}.

\section{Methodology}

Zigbee is a prominent wireless communication standard designed for low-power, low-data-rate applications. It operates based on the IEEE 802.15.4 standard and is widely used in various IoT applications, including smart building systems. 

Zigbee networks are formed using three types of devices: coordinators, routers, and end devices. The coordinator initiates the network and manages its operations, routers extend the network range by forwarding data, and end devices perform specific functions (e.g., sensors, actuators) with minimal power consumption. This hierarchical structure supports robust and scalable network topologies, including star, tree, and mesh configurations. Zigbee's key features are:

\begin{itemize}
    \item Low Power Consumption: Zigbee end devices are designed for long battery life, making them suitable for applications where devices need to operate independently for extended periods.
    \item Short-Range Communication: Zigbee operates in the 2.4 GHz ISM band (with regional variations in frequency bands), supporting communication ranges up to 100 meters, which can be extended through mesh networking.
    \item Low Data Rates: Zigbee supports data rates up to 250 kbps, which is adequate for sensor and control applications but not for high-bandwidth tasks.
    \item Affordability: Zigbee modules and components are generally less expensive compared to other wireless communication technologies like Wi-Fi and Bluetooth.
    \item Non-invasive and Cost-effective Deployment: Zigbee is wireless and the end devices are (usually) battery-driven, therefore no invasive alternations to the building stock have to be done, making it an attractive choice for building owners.
\end{itemize}

Lastly, Zigbee is based on an open standard~\cite{zigbeespec} by the Connectivity Standards Alliance (CSA), formerly, the Zigbee Alliance, which encourages competition among manufacturers and helps keep prices low. Multiple vendors produce Zigbee-compatible devices, leading to a competitive market that drives down costs for consumers.

Analyzing the Zigbee specification reveals several areas where cross-vendor compatibility issues can arise. These issues often stem from the flexibility and options provided within the specification itself, leading to variations in implementation across different manufacturers.

\begin{figure}[htb]
    \centering
    \includegraphics[width=\linewidth]{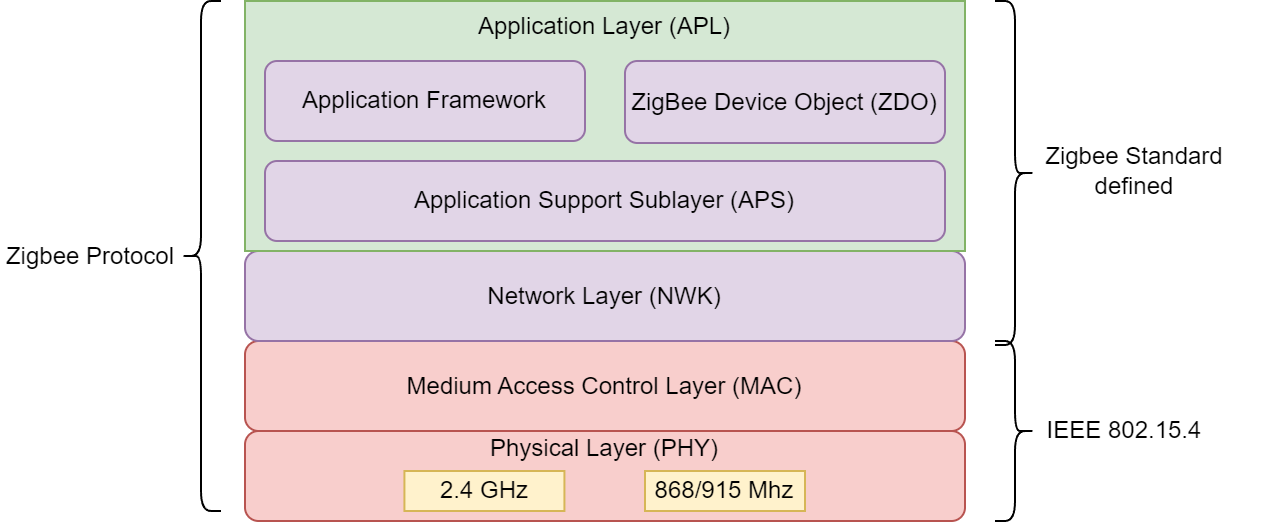}    
    \caption{Zigbee Protocol Stack}
    \label{fig:zigbee-stack}
    
\end{figure}

To understand these issues better, it's essential to delve into the inner workings of the Zigbee stack. The Zigbee stack, as seen on Fig~\ref{fig:zigbee-stack}, is divided into several layers, each responsible for different aspects of the communication process.

Beginning from the bottom; the IEEE 802.15.4 defined physical (PHY) and medium access control (MAC) layers is a foundational technology for low-rate wireless personal area networks (LR-WPANs). It provides the underlying framework for handling the transmission and reception of bits over radio. Moving up a block is the network layer (NWK). The NWK manages network formation, routing, and device addressing. It supports various network topologies, including star, tree, and mesh networks. The next is the application support sublayer (APS), which provides services for binding, group management, and security. On top of the APS, is the Zigbee Device Objects (ZDO) responsible for device discovery, network management, and security management. It ensures that devices can join and leave networks, establish connections, and authenticate securely. The application framework contains the application objects and profiles that define the specific functionalities of Zigbee devices. It includes clusters and attributes that describe device capabilities. Within a Zigbee network, there must be a device running a Trust Center application that manages network and link keys, network access control, and device authentication. Since Zigbee 3.0 all devices are required to have an installation code key to enter and ultimately join a Zigbee network. These install codes are best delivered to the Trust Center "out-of-band"~\cite{zigbeespec,zigbeesecurity} to avoid security compromise. Install code lengths can vary between 6, 8, 12, or 16 bytes (depending on the specification implementation), followed by a 16-bit cyclic redundancy check over those bytes.

The out-of-band delivery of installation codes is crucial for maintaining the security of Zigbee networks. By transmitting installation codes outside the main radio communication channels, the risk of interception and unauthorized access is significantly reduced. However, the lack of a standardized method for this out-of-band delivery has led to inconsistencies across different manufacturers. A common delivery method is a graphical user interface taking a string of characters or a quick-response (QR) code containing said string. However, as innocent as this seems there are many different approaches to the representation of such strings. This variability can cause complications in network setup and integration, especially in environments with devices from multiple vendors. 

Table~\ref{tab:vendor-qrs} showcases different QR code formats\footnote{The \textit{X}'s denote a string of arbitrary length.} from four different Zigbee vendors containing an installation code in hexadecimal needed for device authentication. Highlighted in boldface is the actual installation code. Each vendor uses a distinct format for encoding the installation code. For example, the first code uses multiple fields separated by special characters like \$ and \%, while the second code uses pipes | as delimiters. The third code is a long alphanumeric string without any clear separation, and the fourth code uses a combination of \$ and \% with a different structure altogether. In environments with devices from multiple vendors, the lack of a standardized format complicates the integration process. Network administrators must handle multiple code formats, increasing the chances of errors and reducing overall system efficiency.

\begin{table}[htb]
    \centering
    \caption{QR code strings from different vendors.}
    \resizebox{\textwidth}{!}{
    \begin{tabular}{|c|l|}
    \hline
        Aqara & G\$M:X\$S:X\$D:X\%Z\$A:X\$I:\textbf{DB6DE11643FDA924FE033323F82C54618132}\\
    \hline
        Develco & |X|\textbf{675F67DE359BF9FEB4DF847042AF032824B5}|\\
    \hline
        Bosch & X\textbf{4CAE140FAD7E94FC70E7E8162985D165}\\
    \hline
        Danfoss & G\$M:X\%Z:X\$I:\textbf{E6402113FF0E2CE074B7C069AE35EB03A0D0}\%M:X\\
    \hline
    \end{tabular}}
    \label{tab:vendor-qrs}
\end{table}

Per the Zigbee specification, installation codes are to be used as input to a one-way hash function. The value of this function is then used as the Trust Center link key to authenticate a device to the network. Interestingly, Bosch, in Table~\ref{tab:vendor-qrs}, represents the installation code in the output format of this hash making it unusable for vendors that expect specification-compliant codes.

Other examples of cross-vendor issues often arise when vendors implement the Zigbee specification but maintain closed ecosystems, leading to devices that do not fully interoperate with those from other manufacturers. Here are some examples of such issues:

\begin{itemize}
    \item Philips Hue and Osram Lightify: Both are prominent smart lighting systems that use Zigbee. While both claim Zigbee compliance, they use proprietary clusters and attributes for certain features. As a result, functionalities like color control and specific lighting effects might not work when trying to control Osram bulbs with a Philips Hue hub, and vice versa.
    \item IKEA TRÅDFRI: Although IKEA’s TRÅDFRI system is Zigbee-certified, it initially had issues interoperating with other Zigbee hubs like Philips Hue. Firmware updates have improved compatibility, but there are still occasional issues with advanced functionalities such as color temperature adjustment and scene control.
\end{itemize}

The core of these compatibility issues lies in the balance between adhering to the Zigbee specification and maintaining a competitive edge through proprietary enhancements. While these enhancements can provide better performance or additional features, they often come at the cost of interoperability. To address these issues, we look towards open-source solutions that can handle these vendor-specific edge cases for a more unified and open Zigbee.

\subsection{Zigbee2MQTT}

Zigbee2MQTT~\cite{z2mgithub} is an open-source project that aims to alleviate the use of a proprietary Zigbee gateway via a serial connected adapter and using Message Queuing Telemetry Transport (MQTT) for device control and events. The Zigbee functionality is provided by an underlying module; zigbee-herdsman~\cite{zigbeeherdsman}. It is an open-source Zigbee gateway solution written with a Node.js backend. The zigbee-herdsman module connects, via serial, to the Zigbee adapter, and depending on the adapter's firmware an underlying implementation is used but a common interface is exposed. The list of supported adapter firmware can be seen in Table ~\ref{zigbeefirmware}. The firmware running on an adapter is an implementation of the open Zigbee specification~\cite{csa-zigbee}.

\begin{table}[]
\caption{Supported adapter firmware, vendor, and support status for Zigbee2MQTT~\cite{z2msupportedadapters}.}\label{zigbeefirmware}
\centering
\begin{tabular}{|l|l|l|}
\hline
\textbf{Name}             & \textbf{Vendor}    & \textbf{Status} \\ \hline
Z-Stack                   & Texas Instruments  & Recommended     \\
deCONZ                    & dresden elektronik & Recommended     \\
ZiGate                    & ZiGate             & Experimental    \\
EmberZNet Serial Protocol & Silicon Labs       & Experimental    \\ \hline
\end{tabular}
\end{table}

To parse messages from and to specific end devices the zigbee-herdsman-converters~\cite{zigbeeherdsmancovnerter} module is used that contains specific device handlers for all types of vendors like IKEA, Siemens, Bosch, etc. Out of the box, it supports more than 3000 devices from over 400 vendors. The support for new devices keeps expanding by community-driven development. Supporting a new device is trivial in the provided web app and can be done during runtime. Finally, Zigbee2MQTT itself utilizes the aforementioned modules and creates MQTT messages from the Zigbee events. Furthermore, it provides a web application for device overview and devices.

MQTT is a lightweight, publish-subscribe network protocol designed for constrained devices and low-bandwidth, high-latency, or unreliable networks. Its design makes it ideal for IoT applications, where efficient, reliable, and real-time communication between devices is crucial. Zigbee2MQTT bridges the gap between Zigbee devices and MQTT, making it easier for users to integrate and manage Zigbee devices through an MQTT-based system without needing in-depth knowledge of Zigbee.

\subsection{Large-scale IoT Architecture for Smart Buildings}

\begin{figure}[htb]
    \centering
    \includegraphics[width=0.9\linewidth]{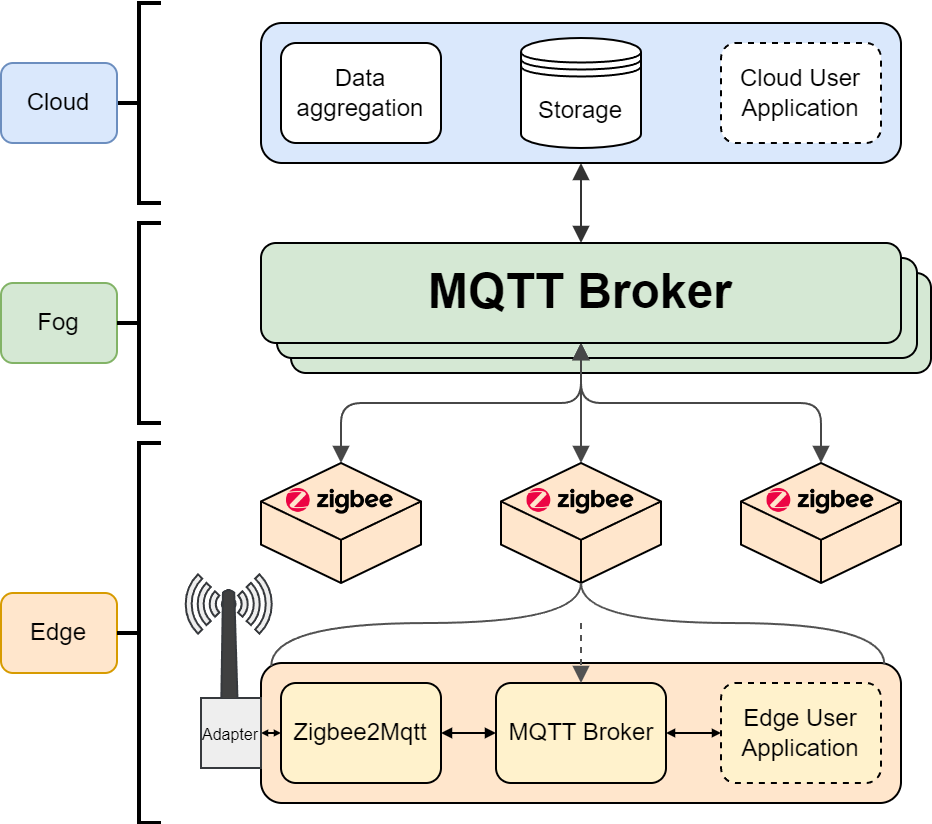}
    \caption{Full IoT Architecture by processing layer.}
    \label{fig:z2m-architecture}
\end{figure}

The provided Figure~\ref{fig:z2m-architecture} illustrates the integration architecture of Zigbee2MQTT, emphasizing its role in bridging Zigbee devices with an MQTT broker for smart building applications. At the edge of the network, a Zigbee adapter connects to a device running Zigbee2MQTT, facilitating communication between the Zigbee network and the Zigbee2MQTT software. This open-source application translates Zigbee messages from the devices into MQTT messages, abstracting the complexities of Zigbee and presenting a unified MQTT interface. Users interact with the MQTT protocol without needing in-depth knowledge of Zigbee, simplifying integration and management.

\newpage

A crucial component in this architecture is the local MQTT broker situated at the edge. This local broker not only manages real-time communication between Zigbee2MQTT and edge applications but also provides offline buffering. This ensures that if there is a network disruption, the local broker can store messages temporarily and relay them once connectivity is restored, maintaining the integrity and continuity of the system. Additionally, the local broker allows user applications to run on the edge, processing MQTT messages locally. This capability supports low-latency responses and local decision-making, crucial for applications that require immediate actions, such as security systems or environmental controls.

Central to this architecture is the MQTT broker, which acts as an intermediary for all communications. Zigbee2MQTT publishes messages from Zigbee devices to the broker, and any application that subscribes to the relevant topics can receive these messages. The MQTT broker manages message routing, ensuring that messages from Zigbee devices are delivered to the appropriate subscribers.

Additionally, the architecture leverages MQTT bridge features to transmit messages from the edge to the fog and cloud layers. This relay mechanism allows data to be efficiently sent to higher layers for aggregation, processing, and long-term storage. The fog layer handles intermediate processing, ensuring that data is appropriately filtered and processed before reaching the cloud.

The cloud layer handles data aggregation and storage, enabling long-term data retention, analysis, and historical querying. Cloud applications can subscribe to MQTT topics to receive real-time data, perform complex analyses, and control Zigbee devices remotely, leveraging cloud computing resources for advanced features such as machine learning, predictive maintenance, and large-scale analytics.

This architecture highlights several key advantages. It facilitates seamless integration by converting Zigbee events into MQTT messages, removing the need for users to handle Zigbee-specific protocols. It is highly scalable, allowing for additional Zigbee devices without significant system changes, and the MQTT broker efficiently manages communications, supporting numerous devices and applications simultaneously. The publish-subscribe model of MQTT offers great flexibility, enabling applications to dynamically subscribe to topics and respond to events in real-time. This decouples Zigbee devices from the applications that use their data, ensuring that changes or upgrades to one part of the system do not necessarily impact other parts.

For example, in a smart building with various Zigbee-based sensors and actuators, Zigbee2MQTT converts data from these devices into MQTT messages. An edge application can use this data for immediate adjustments, such as turning on lights when motion is detected, while cloud applications analyze long-term trends to optimize HVAC settings for energy efficiency. By leveraging MQTT, this architecture ensures that the smart building system is robust, scalable, and easy to manage, promoting efficient resource use and improving the overall user experience.

\section{Case Study}

Zigbee2MQTT itself can run on most operating systems, but in this case study it was executed in a Docker environment on an Ubuntu Desktop 22.04 host. The only necessary change is to expose the serial adapter from \texttt{/dev/tty*} into the container. To unlock the ability to utilize Zigbee any supported adapters~\cite{z2msupportedadapters} can be used. This paper conducted its research with the USB ZBDongle-E from Sonoff running the experimental EmberZNet Serial Protocol (EZSP). An adapter can run either as a coordinator or router depending on the use case, but in this case study one coordinator and no routers were used.

\begin{figure}[htb]
    \centering
    \includegraphics[width=0.70\linewidth]{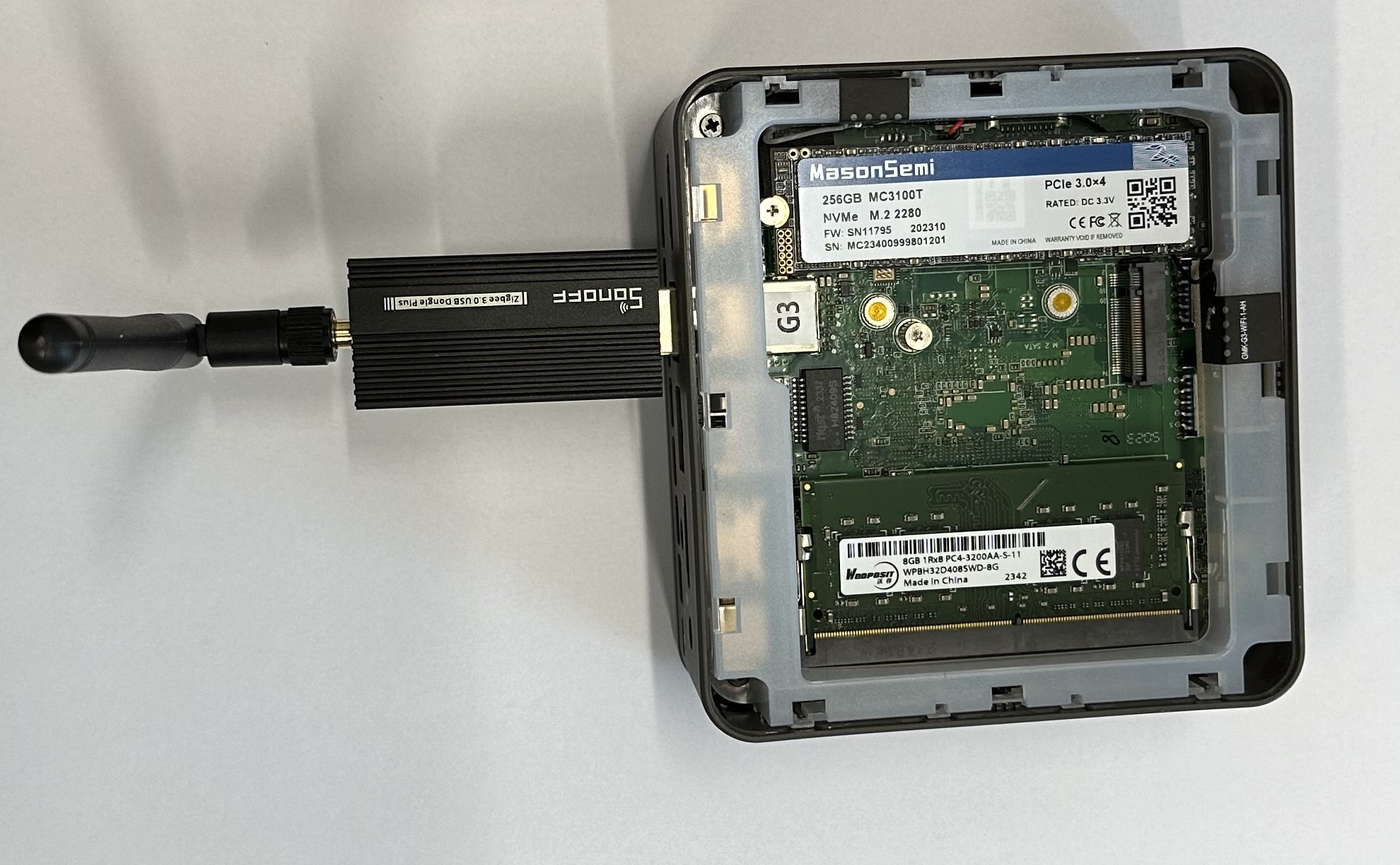}
    \caption{Mini-PC and Zigbee adapter.}
    \label{fig:mini-pc}
\end{figure}

Figure~\ref{fig:mini-pc} shows the example gateway (combined mini-PC and Zigbee adapter) setup used in this case study. The mini-PC sports 8GB of DDR4 RAM, Intel N100 4 cores CPU, and 256GB of NVMe storage for a cost of 120€. The cost of the Sonoff USB ZBDongle-E is 25€. Total cost of the gateway is 145€.

\begin{figure}[htb]
    \centering
    \includegraphics[width=0.74\textwidth]{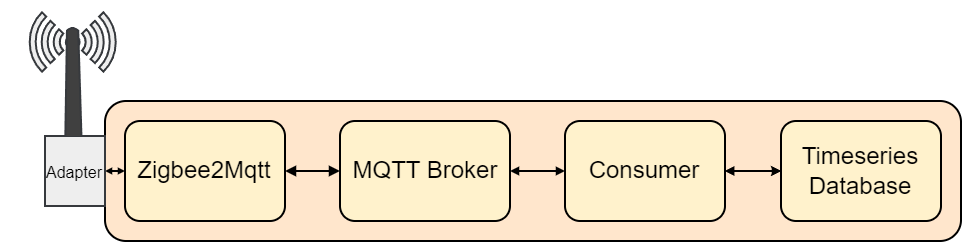}
    \caption{Case Study Gateway Architecture.}
    \label{fig:minipc-architecture}
\end{figure}

To test the edge computing capabilities, the case study was conducted with the proposed architecture running locally on the gateway, as seen on Fig~\ref{fig:minipc-architecture}. The consumer component subscribes to data from the gateway's sensors and stores them in a time-series database.

\begin{figure}[htb]
    \centering
    \includegraphics[width=0.8\textwidth]{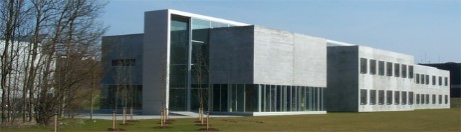}
    \caption{Maersk building.}
    \label{fig:maersk}
\end{figure}

The case study took place in two offices on the second floor in the Mærsk Mc-Kinney Møller building on the University of Southern Denmark's campus illustrated on Fig~\ref{fig:maersk}. The offices are separated by a catwalk and two concrete walls.

\begin{figure}[htb]
    \centering
    \includegraphics[width=0.53\textwidth]{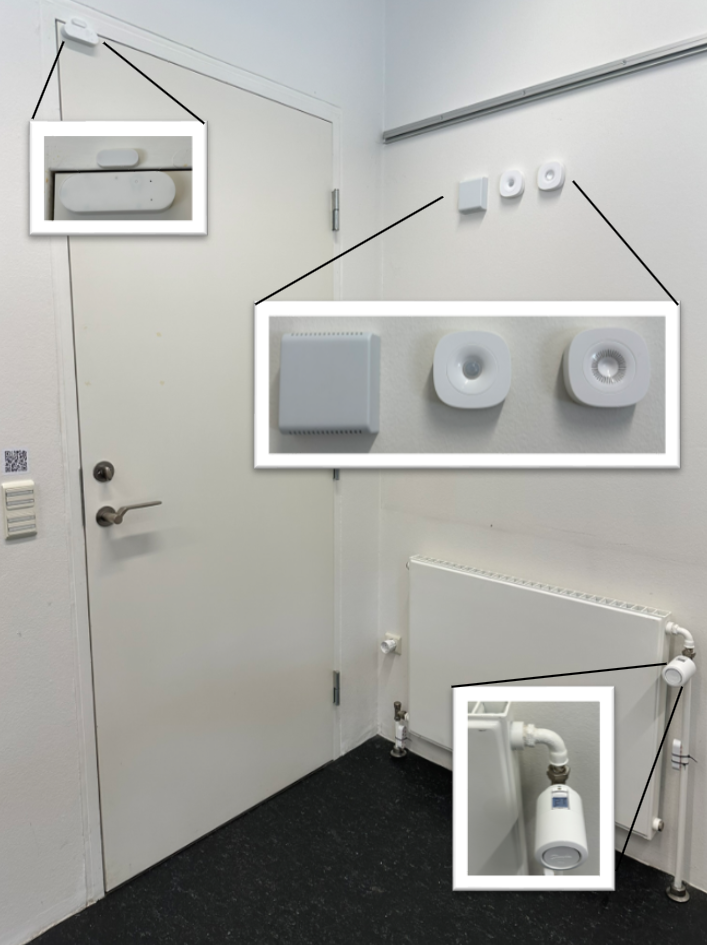}
    \caption{Office sensor deployment.}
    \label{fig:config}
\end{figure}

One of the office's sensor deployments can be seen on Figure \ref{fig:config}. A total of 10 sensors were installed, including 2 smart thermostats, 2 air quality sensors, 2 contact sensors, 2 motion sensors, and 2 CO2 sensors. The installation process was simply plugging in the battery in the sensors while the Zigbee2Mqtt gateway was in pairing mode.

\newpage
\section{Results}

\begin{figure}[!htb]
    \centering
\includegraphics[width=0.85\textwidth]{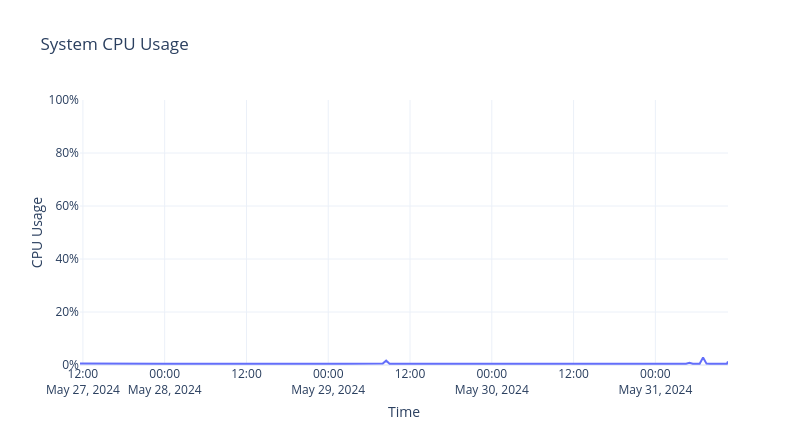}
    \caption{System CPU Usage}
    \label{fig:system-cpu}
\end{figure}

The results section presents various graphs that illustrate the performance and efficiency of the proposed edge computing gateway setup for smart buildings. Figure~\ref{fig:system-cpu} indicates minimal CPU usage, demonstrating the efficiency of the mini-PC setup in handling multiple sensors with low resource consumption. This efficiency further underscores the system's suitability for scalable, cost-effective smart building management.

\begin{figure}
    \centering
    \includegraphics[width=0.85\textwidth]{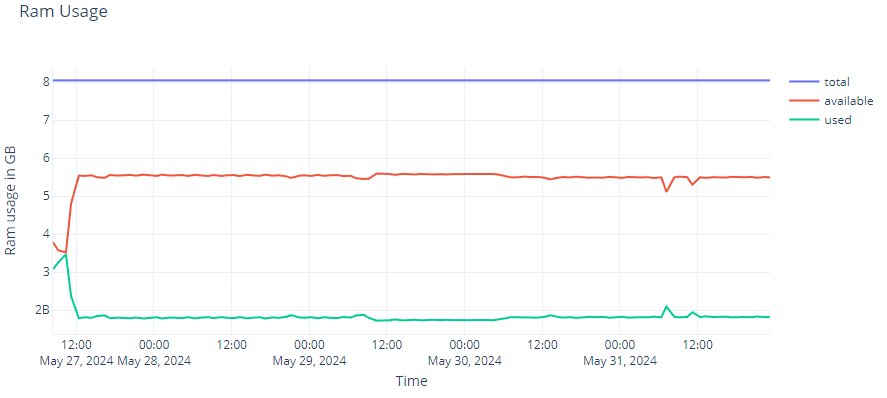}
    \caption{System RAM usage}
    \label{fig:system-ram}
\end{figure}

 Similarly, Figure~\ref{fig:system-ram} highlights the system RAM usage, showing that the mini-PC setup has huge edge computing potential with the amount of available RAM.

\begin{figure}[!htb]
    \centering
    \includegraphics[width=0.85\textwidth]{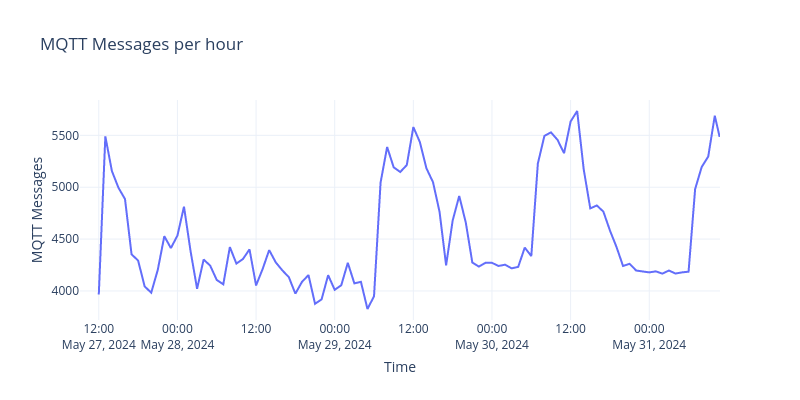}
    \caption{System MQTT messages per hour.}
    \label{fig:system-mqtt}
\end{figure}

Figure~\ref{fig:system-mqtt} shows the MQTT messages per hour, where on the 28th and 29th there was low occupant traffic in the offices, but normal office hours were observed on the 30th and 31st. This graph illustrates the variation in message traffic based on occupancy.

\begin{figure}
    \centering
    \includegraphics[width=0.85\textwidth]{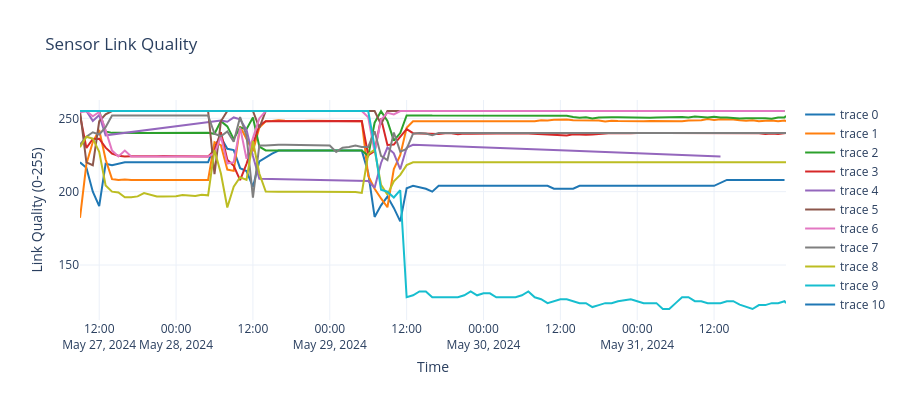}
    \caption{Sensor link quality.}
    \label{fig:system-linkquality}
\end{figure}

Figure~\ref{fig:system-linkquality} presents the sensor link quality, reflecting the stability and reliability of the wireless connections within the smart building setup. High link quality indicates that the Zigbee network maintains strong and consistent communication links, which is crucial for ensuring the dependable operation of the deployed sensors. Note that trace 9's dip in link quality is due to the sensor being moved to another office.

This architecture showcases a practical, efficient, and cost-effective way to manage and utilize Zigbee devices within a smart building setup. By leveraging edge computing and MQTT, the system benefits from a flexible and scalable communication model, ensuring reliable and real-time data processing and integration. Furthermore, leveraging the open-source Zigbee2MQTT allows for frictionless cross-vendor compatibility.

\section{Discussion}

One of the ongoing challenges in the Zigbee ecosystem is balancing the freedom of device choice with the constraints imposed by proprietary implementations of the Zigbee protocol. While the Zigbee standard aims to facilitate interoperability across different devices and manufacturers, proprietary extensions and custom implementations can limit this interoperability.

Moreover, the hardware utilized in the case study is modifiable and can be adapted to meet specific edge computing requirements. This flexibility enables further cost reductions and customization based on the particular needs of the smart building deployment. By tailoring the hardware configuration, it is possible to optimize performance and cost-effectiveness, ensuring that the system remains efficient and scalable even as the demands of the smart building evolve. For instance, the mini-PC and Zigbee adapter employed in the case study, with a total cost of 145€, can be adjusted or upgraded depending on specific processing needs or budget constraints, thereby enhancing its cost-effectiveness.

The introduction of Zigbee PRO 2023~\cite{zigbeepro} aims to address some of these interoperability challenges by further enhancing the Zigbee standard. Zigbee PRO 2023 includes improvements in security, network scalability, and device interoperability. The certification process ensures that devices meet stringent requirements for performance and interoperability. However, the cost of certification can be a barrier for some manufacturers, potentially limiting the variety of certified devices available on the market. 

Looking forward, the Matter~\cite{matter} protocol, developed by the CSA and open source contributors, promises to further enhance compatibility and interoperability across smart home devices. Matter aims to create a unified IP-based connectivity standard that works across different ecosystems, including Zigbee, Z-Wave, and Wi-Fi. By adopting Matter, future Zigbee devices can achieve even greater interoperability, ensuring seamless integration and operation within mixed-technology environments. This move towards a universal standard is expected to simplify device choice for consumers and reduce the fragmentation currently seen in the market.

The integration of eSIM (embedded SIM)~\cite{esim} and cellular technologies, such as NB-IoT (Narrowband IoT)~\cite{nbiot}, presents additional opportunities for enhancing smart building systems. These technologies offer reliable and scalable connectivity options, especially in scenarios where traditional Wi-Fi or Zigbee networks may not be feasible. NB-IoT, for instance, provides wide-area coverage with low power consumption, making it ideal for deploying sensors and devices in remote or hard-to-reach areas.

Lastly,  while this study provides valuable insights into the development of a cost-effective, scalable, and open-source Zigbee gateway for smart buildings, it does have some limitations. The cost analysis presented focuses on the Zigbee technology, with limited exploration of other wireless protocols. A more thorough cost analysis, including a comparison of Zigbee with alternative technologies such as Wi-Fi, BLE, and Z-Wave, is needed to fully understand the economic implications of different choices. Additionally, future work should incorporate more detailed benchmarks that compare performance, scalability, and energy efficiency across these technologies in varied deployment scenarios. Such an analysis would provide a clearer understanding of the trade-offs and help stakeholders make more informed decisions when selecting technologies for smart building retrofits.
 
\section{Conclusion}

This study introduced a novel, cost-effective, scalable, and open-source Zigbee gateway designed for smart building applications. By leveraging open-source tools and low-cost hardware, we have significantly lowered the economic barriers for upgrading existing buildings to smart buildings. The proposed architecture bridges Zigbee devices with an MQTT broker, enabling seamless integration and real-time management through a unified and user-friendly interface.

Despite the advantages of Zigbee, compatibility concerns persist due to proprietary enhancements and variations in implementation across different manufacturers. These challenges often lead to interoperability issues when integrating devices from multiple vendors. Our approach using Zigbee2MQTT addresses these concerns by providing a unified MQTT interface, mitigating issues related to vendor lock-in.

The case study demonstrated the practical application and edge computing capabilities of the proposed gateway. By running Zigbee2MQTT in a Docker environment on an Ubuntu host, we showcased the system's efficiency in handling multiple sensors with minimal resource consumption. The local MQTT broker provided offline buffering and enabled low-latency responses, ensuring reliable and real-time data processing.

In conclusion, the introduced architecture and open-source Zigbee gateway present a robust solution for enhancing the efficiency, reliability, and affordability of smart buildings.

\subsubsection{Acknowledgements.} This paper is part of the project “CELSIUS - Cost-effective large-scale IoT solutions for energy efficient medium- and large-sized buildings”, funded by the Danish funding agency, the Danish Energy Technology Development and Demonstration (EUDP) program, Denmark (Case no. 64020-2108)

\subsubsection{Disclosures.} The authors have no competing interests to declare that are relevant to the content of this article.

%
%
%
%

\end{document}